\documentclass{ametsoc}

\journal{jcli}

\bibpunct{(}{)}{;}{a}{}{,}

\title{Impact of anthropogenic climate change on the East Asian summer monsoon}

\authors{Claire Burke\correspondingauthor{Met Office Hadley Centre, Fitzroy Road, Exeter, EX1 3PB, UK.} and Peter Stott}

\affiliation{Met Office Hadley Centre, UK}

\email{claire.burke@metoffice.gov.uk}

\abstract{The East Asian summer monsoon (EASM) is important for bringing rainfall to large areas of China. Historically, variations in the EASM have had major impacts including flooding and drought. We present an analysis of the impact of anthropogenic climate change on EASM rainfall in Eastern China using a newly updated attribution system. Our results suggest that anthropogenic climate change has led to an overall decrease in total monsoon rainfall over the past 65 years, and an increased number of dry days. However the model also predicts that anthropogenic forcings have caused the most extreme heavy rainfall events to become shorter in duration and more intense. With the potential for future changes in aerosol and greenhouse gas emissions, historical trends in monsoon rainfall may not be indicative of future changes, although extreme rainfall is projected to increase over East Asia with continued warming in the region.}

\begin{document}

\maketitle

%








\section{Introduction}
The East Asian summer monsoon (EASM) brings much needed water for agriculture to most of Eastern China. In recent decades southern provinces of China have experienced an increased frequency of severe flooding during the monsoon season. In contrast northern provinces of China have experienced an increase in severe summer droughts (for details of the northern drought / southern flood pattern see, for example, \cite{QianZhou14}, also see FloodList Copernicus project for examples, http://floodlist.com/tag/china). Understanding changes in past and future monsoon rainfall patterns can have important implications for water management and urban planning.

The Clausius-Clapeyron relation states that the atmosphere can hold 7\% more moisture per degree of warming. Basic physical expectations are that a warmer world should experience increased amounts of rainfall. A simple interpretation of the Clausius-Clapeyron relation is that the total quantity of rainfall should increase by 7\% per degree of warming globally. However, in reality different surfaces heat at different rates, and in the case of anthropogenically induced global warming, greenhouse gases do not cause the atmosphere to be heated equally at all levels. Additionally, the emission of aerosols can change cloud formation properties, alter the locations of cloud nucleation sites, and cause localised cooling. Changes in chemistry and thermodynamics mean that increases in temperature may not necessarily lead to a uniform increase in precipitation in all locations or at all intensities of rainfall.

Heating of the lower troposphere as a consequence of increased concentrations of well mixed greenhouses gases (GHGs) leads to an increase in the height of the tropopause. GCM-based studies have argued that warming will cause increases in cloud height and stronger convection as a result \citep[see][]{Fowler95, Mitchell92, Trenbeth2003}. Other studies have argued that surface warming leads to decreases in convective mass fluxes with the heating of the upper troposphere. It is instead argued that increases in horizontal transport due to an enhanced pattern of evaporation minus precipitation will cause increased convergence \citep[e.g.][]{HeldSoden2006}. With an increased moisture content of the air, stronger convection or convergence will lead to more severe storms with higher hourly and total rainfall \citep{Fowler95,Trenbeth2003,HeldSoden2006}. 

Several studies of global rainfall trends have found that global annual mean and total precipitation has increased by 1--3\% per degree of warming \citep[e.g.,][]{AllenIngram2002,Wu13,Donat16}. At the same time extreme rainfall, defined by upper decile daily total or Rx1day, has increased by 6--7\% per degree of warming \citep[e.g.,][]{Trenbeth2003,Westra13}. The increase in extreme heavy rain is often found to be at the expense of light rain, with studies finding a decrease in the number of light rain days, or total rain from light rain events, coinciding with increased totals or frequency of heavy rain \citep[e.g.,][]{Trenbeth2003,Ban15,AllenIngram2002}. 

Regional changes in rainfall totals and a changing distribution of rainfall between light and heavy events are also observed. In the current study we focus our attention on China. The annual rainfall climatology of China can be broadly split into two halves, a cold, dry winter monsoon from October to March, and a warm, wet summer monsoon from April to September. During the winter monsoon continental cold, dry air flows southwards from high latitudes, bringing a cold, dry winter. During the summer monsoon, warm moist air flows from the ocean to the south of China and converges with the cool dry air to the north. The convergence causes the formation of a rain band over the Indochina peninsula in China, and as the summer season progresses the rain band moves steadily northwards over Eastern China (and is known as the Meiyu), eventually as far north as Japan (where it is referred to as the Baiyu) and Korea (where it is referred to as the  Changma). Towards the end of the summer the rain band then retreats southwards \citep[for a summary of the characteristics of the East Asian summer monsoon see][]{YihuiChan05, Hsu14,Feng15}. 

As mentioned above, in recent years Southern China has seen more frequent incidents of flooding and Northern China has seen more frequent severe droughts during the monsoon season when compared to historical monsoon seasons. Changes in monsoon total rain, and changes in circulation patterns which dictate the most northern extent of the Meiyu front each year have been correlated with modes of natural variability, such as the Pacific Decadal Oscillation (PDO) \citep{Zhu2011,QianZhou14,Qian14}. Several studies have also noted changes in total summer rain which coincide with the increasing trend in global temperature \citep{Liu05,Zhai04,Su06,Fu08}, and some studies suggest links with local emissions of anthropogenic aerosols \citep[e.g.][]{Qian2009,FuDan13,DengXu15}. Many studies also note a change in character of summer rainfall in Eastern China, with increases in numbers of heavy rain days and decreases in numbers of light rain days reported \citep{Zhai04,Liu05,FuDan13,Fu08}. 

In this study we examine changes in the East Asian Summer Monsoon (EASM) rainfall over China using an ensemble of simulations from an atmosphere-only climate model representing present-day conditions with anthropogenic influences, and comparing these to an ensemble representing conditions without anthropogenic influences. We compare characteristics of light and heavy rain during the monsoon in model experiments with and without climate change and compare our results with those of previous observational studies.

\section{Data}

We use a model ensemble from HadGEM3-A-N216, run in the atmosphere-only mode with prescribed historical sea surface temperatures (SSTs) from HadISST1 \citep{hadISST}. The resolution is approximately 0.5 x 0.8 degrees, equivalent to $\sim$50 km at the latitude range covered by China. The ensemble contains 15 members which include both anthropogenic and natural forcings (denoted ALL) from 1960-2015. 
This is compared with an ensemble of 15 runs of the same model which contain only natural forcings (denoted NAT), in which the SSTs have been adjusted to remove anthropogenic warming. This anthropogenic warming is calculated from the difference between the mean patterns derived from ALL and NAT simulations in 19 model ensembles from CMIP5 \citep{CMIP5}. This pattern of SSTs is subtracted off the SSTs for the ALL experiment to provide the SSTs used in the NAT experiment. We also adjust the sea-ice concentration for the NAT experiment using simple empirical relationships between SSTs and sea-ice concentrations. These methods and full details on model experiment setups are described in \cite{Christidis2013} and Ciavarella et al (2017, in prep).

To verify the model output we use the APHRODITE observational gridded daily precipitation dataset for East Asia \citep{Yatagai2012}. This dataset runs from 1960-2007, and is gridded to approximately the same resolution as the model (0.5x0.5 degrees).
\cite{HanZhou12} compare the APHRODITE dataset to daily rainfall records from 559 rain gauges spread over China. They find that the APHRODITE data shows very similar rainfall amounts for mean variables, such as seasonal total, and accurately characterises the progression of the seasonal rain band. However they find that the gridding of spatialy sparse station data in APHRODITE leads to underestimates of precipitation intensity and overestimates of precipitation frequency compared to the station data. They show that annual mean heavy rainfall totals are underestimated and light to moderate rainfall totals are overestimated in the gridded data. A large difference is also found between the station and APHRODITE data for spatial patterns of trends in intense rainfall, and that the APHRODITE data underestimate trends in the recent northern drought / southern flood pattern compared to station data. With these limitations in mind, we use the APHRODITE data for model verification of seasonal rainfall characteristics, and focus on the model output for examining trends in rainfall and changes in extreme rainfall characteristics.

For consistent comparison, we regrid both the observations and model to an identical 1x1 degree grid, taking daily area means over the cells within the 1x1 degree grid. A map of the mean and maximum numbers of stations per grid cell for APHRODITE between 1960 and 2007 is shown in the top row of Figure~\ref{obs_stations}. As is clear in the figure, in Western China station coverage is spatially very sparse. Additionally, being a desert, the monsoon does not reach this region, so we exclude Western China from our analysis.

\section{Model evaluation and climatology}

For this study we define the monsoon season to be from the beginning of April to the end of August.
Figure~\ref{pent_clim} shows the climatological rainfall, averaged over 1960-2000, for the monsoon season from 5-day total rainfall for 4 time slices throughout the monsoon season. Being a multi-decadal average the detailed features of the monsoon do not appear very strongly due to their spatial variation between years. However some indication of the general location of the Meiyu front can be seen in both observations and model. The model reproduces fairly well the spatial location of the observed rainfall and the progression of the locations of high and low rainfall throughout the monsoon season. However the model consistently overestimates the total rainfall. When normalised to the observations (dividing out by the East China area-mean ratio of observed total to model total, right hand column in Figure~\ref{pent_clim}) the model appears qualitatively similar to the observed rainfall patterns.

Figure~\ref{pent_clim} also shows the climatological (1960--2000) mean total seasonal rainfall and climatological seasonal maximum daily rainfall. As is again clear in this figure, the model reproduces quite well the spatial patterns of rainfall but tends to over-predict rainfall totals. When normalized the model mean appears qualitatively similar spatially to the observations. We use the raw (non-normalized) model output for the rest of our evaluation and for our analysis of the monsoon.

\begin{figure}
\centering
\includegraphics[width=15cm]{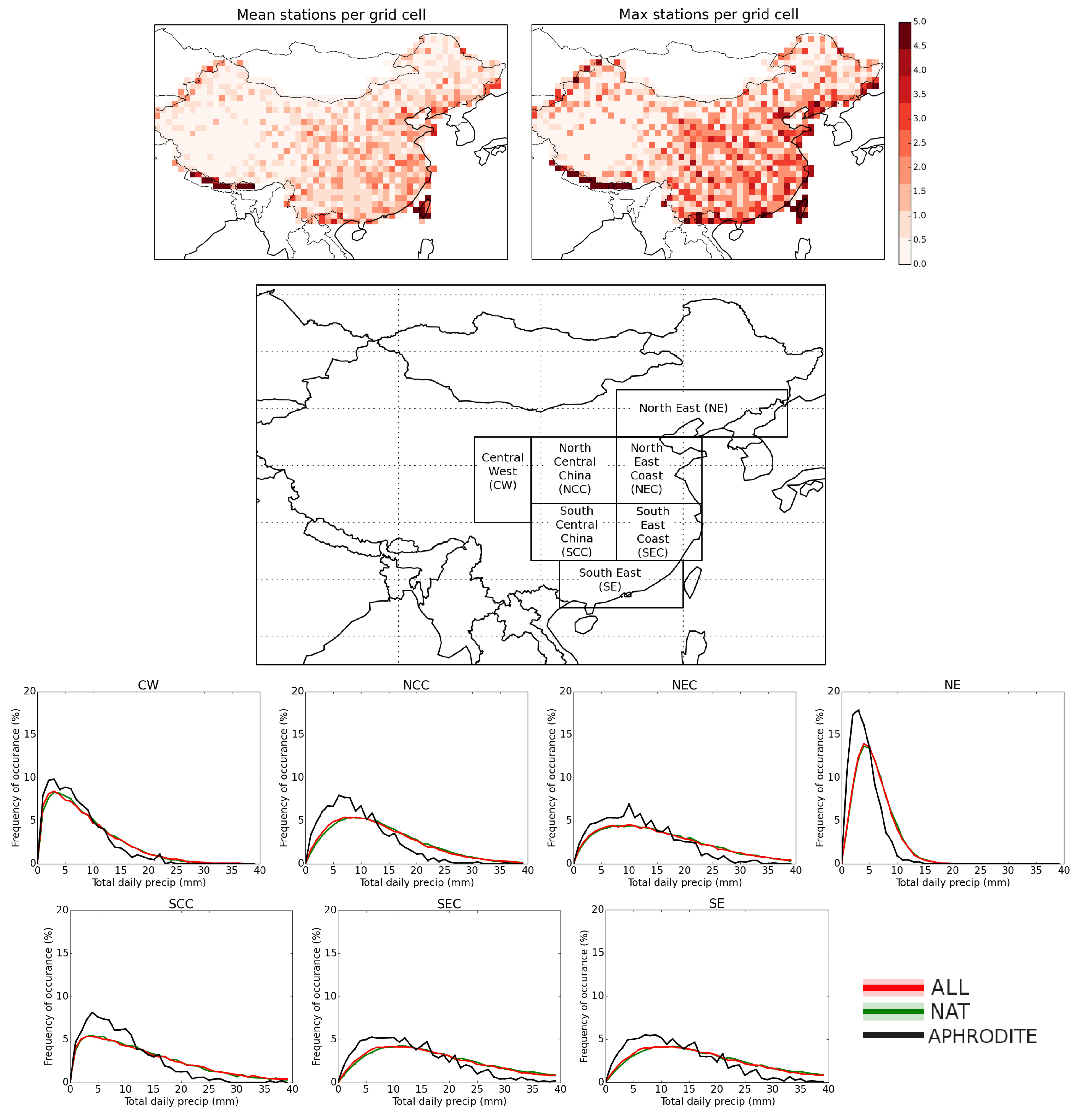}
\caption{Top row: 1960--2007 mean (left) and maximum (right) number of sations per square-degree grid cell from which the APHRODITE obervation data is constructed. Centre: China divided into climatologically similar regions. For verification we exclude areas of China with very low observation station density and very little total monsoon rainfall. Bottom rows: Precipitation intensity distribution (from area daily mean) for regions in China, climatology for 1960-2000 - the contribution of daily rainfall total to the total monsoon rainfall, black line is observations, red and green are ALL and NAT experiments respectively (the green line is often hidden behind the red in these plots).}
\label{obs_stations}
\end{figure}

\begin{figure}
\centering
\includegraphics[width=15cm]{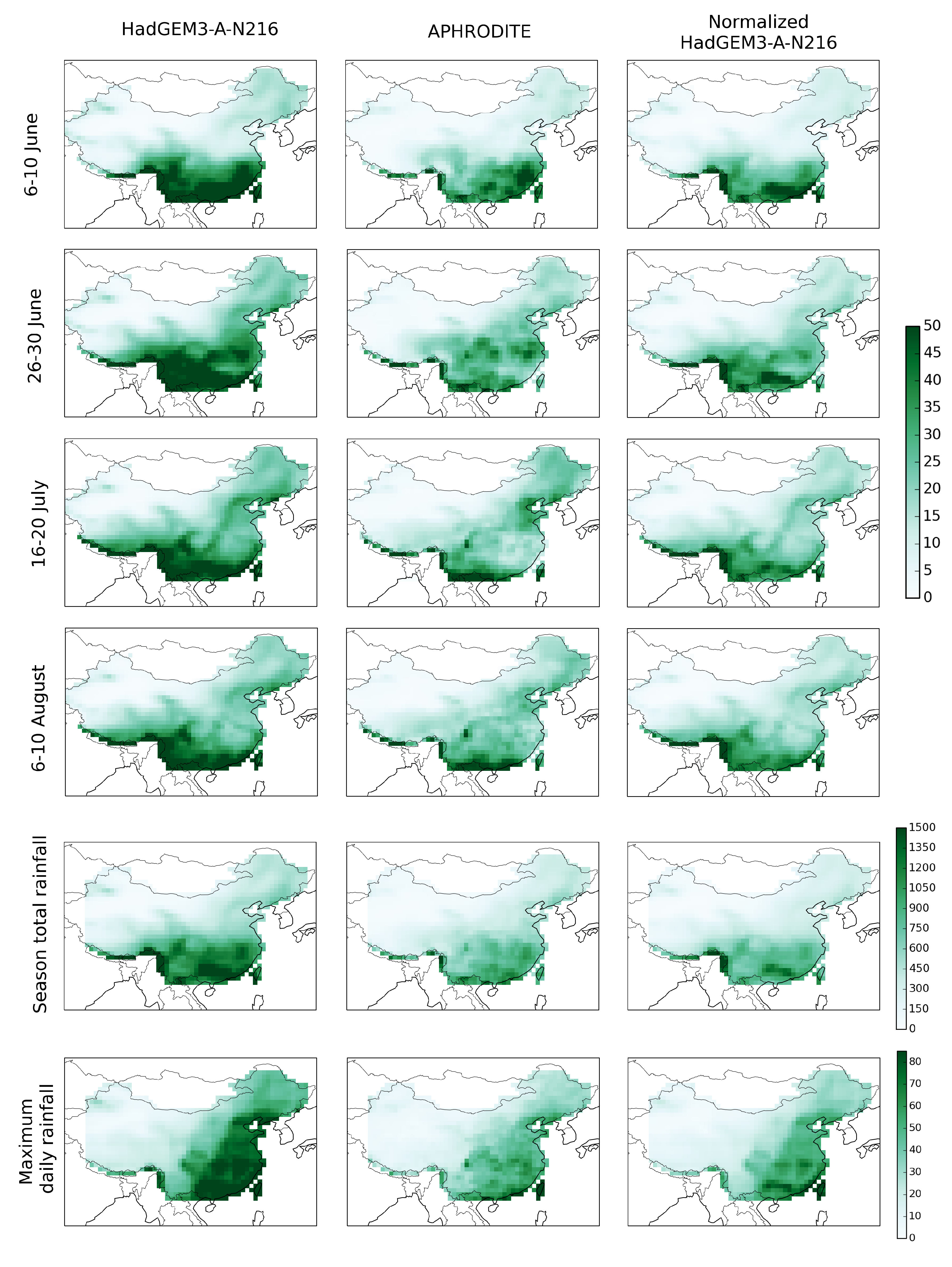}
\caption{Top 4 rows: 1960--2000 mean pentad climatology (5-day total rainfall, mm/5 days) for observations (centre) and ALL-forcing ensemble mean (left), and model mean when normalized to the observed average (right). Second from bottom: Seasonal mean total rain (mm) for monsoon season for mean of 1960-2000. Bottom row: Maximum daily total rain (mm) for monsoon season, mean of 1960-2000.}
\label{pent_clim}
\end{figure}

We group areas of China into climatologically similar regions, indicated in Figure~\ref{obs_stations}. We exclude regions with very low numbers of observation stations. These regions also tend to be in the desert parts of China and therefore receive very little rainfall annually and are not climatologically subject to rainfall as a result of the monsoon.

The bottom rows of Figure~\ref{obs_stations} show the intensity distribution of daily precipitation total for all years between 1960-2000. This figure indicates how much daily total rainfall contributes to the total monsoon seasonal rain. For the central 4 regions the model reproduces the shape of the distribution well. However for all the regions the model peak of the distribution of daily rainfall contribution is at a somewhat larger value than is observed, and shows a fatter tail at the high daily total end of the distribution. However, as previously noted, the APHRODITE gridded data may underestimate the heavier end of the daily precipitation distribution. This could lead to a skewing to the lighter end of daily precipitation in the observations. Alternatively it could be that the model systematically overestimates daily rainfall in Eastern China during the monsoon season.

\begin{figure}
\centering
\includegraphics[width=15cm]{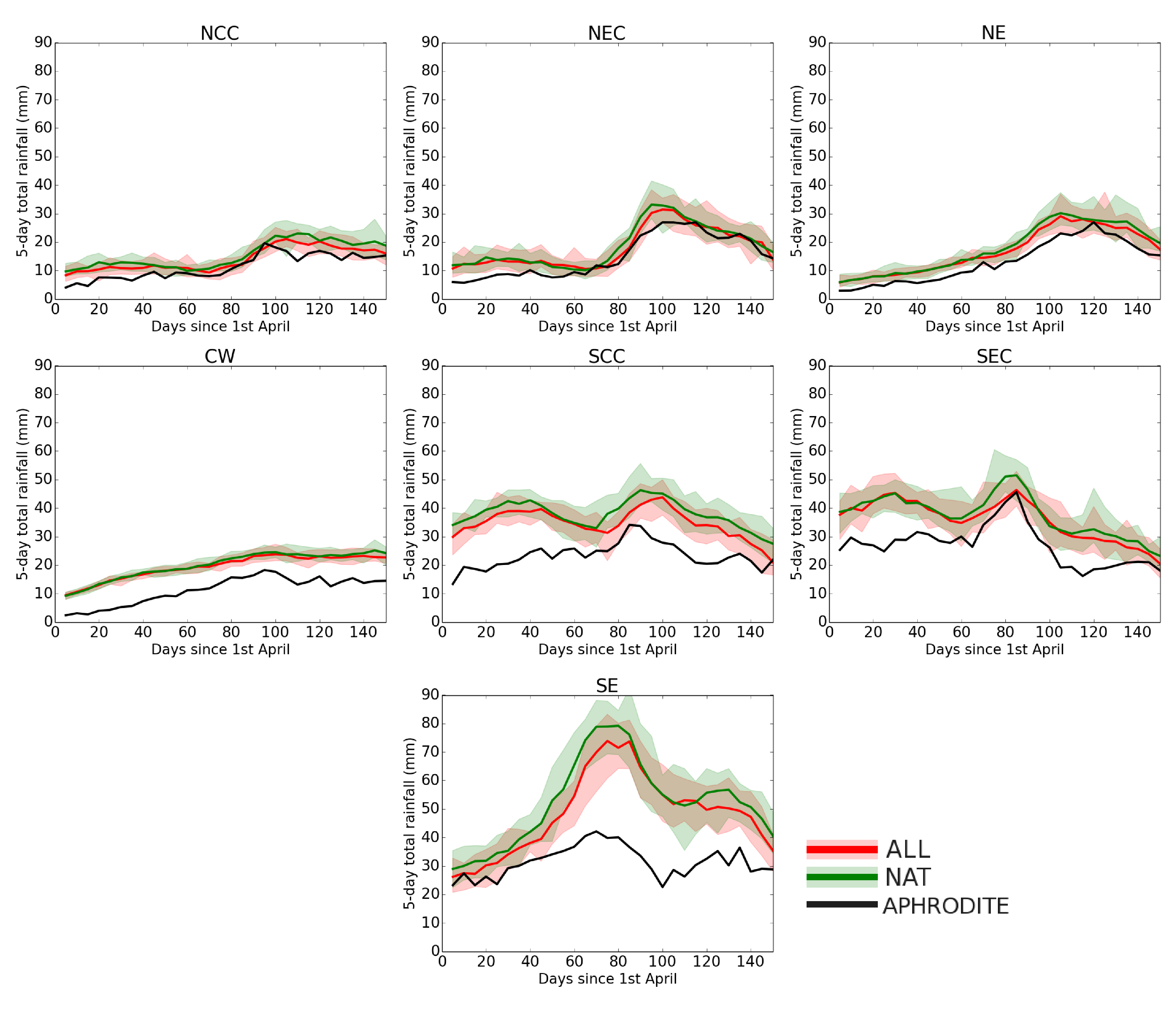}
\caption{5-day total rainfall time series throughout the monsoon season, averaged over 1960--2000. Regions corresponding to Figure~\ref{obs_stations} are indicated above panels. Red and green lines are ALL-forcings and NAT-forcings ensemble means, red and green shading are ensemble range (appears brown where the two overlap). Black line is the observations. }
\label{mons_shape_fig}
\end{figure}

Figure~\ref{mons_shape_fig} shows the 1960-2000 climatology of 5-day consecutive (non-overlapping) total rain throughout the monsoon season for the regions shown in Figure~\ref{obs_stations}. As in earlier figures, the model reproduces the spatial patterns and timing of the monsoon rainfall fairly well but overestimates the total rainfall. For three northern regions the model spread encompasses the observed totals. For SEC the model spread and mean are close to the observed values but generally for the southern regions the mean 5-day totals are greater than observed. 

\begin{figure}
\centering
\includegraphics[width=15cm]{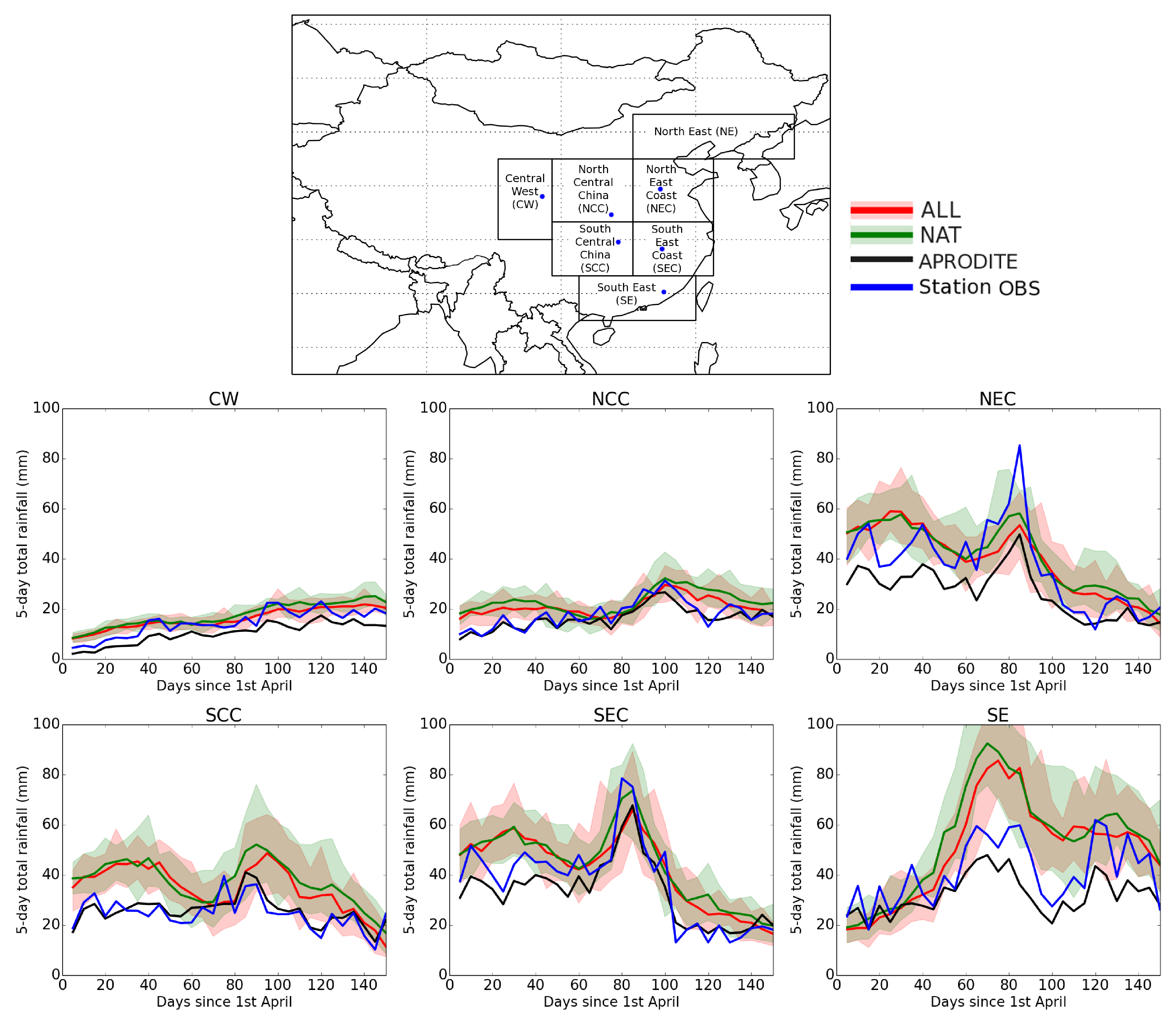}
\caption{5-day total rainfall time series throughout the monsoon season, averaged over 1960--2000. Blue line shows data for an individual station, indicated in the map with a blue dot. Black line is APHRODITE.  Red and green lines are ALL-forcings and NAT-forcings ensemble means. The APHRODITE, ALL and NAT are for the individual grid cell in which the station lies.}
\label{mons_shape_ver}
\end{figure}

The reported under-estimation of extreme rainfall in APHRODITE \citep{HanZhou12} may contribute to the discrepancy between observations and models. 
We also examine this claim using a small number of publicly available station data for China, which have undergone basic quality control. In Figure~\ref{mons_shape_ver} we show the same as Figure~\ref{mons_shape_fig} but for one station per region for 6 of 7 regions, compared with grid cells containing the station location in the model and APHRODITE data - station locations are indicated in the figure. Whilst the station is a point source, and the gridded data is a representation of a larger area this comparison gives a reasonable idea of how well the model and gridded observations perform. In Figure~\ref{mons_shape_ver} it is generally clear that the station data 5-day rainfall totals are slightly higher than the APHRODITE data. As noted in \cite{HanZhou12}, in Figure~\ref{mons_shape_ver} the APHRODITE data shows notably lower total rain for heavy rainfall days than is recorded in the station data. This figure shows the model data to be more similar to the station data than the APHRODITE data. This comparison provides some crude measure of observational uncertainty. While the station data is a point source, estimates of 5 day total rainfall may be less biased than the larger grid box average from APHRODITE.

When compared to APHRODITE our model reproduces the main features of the monsoon fairly accurately. Comparison with data from a few stations shows that the model also reproduces extreme rainfall. Although the model seems to generally overestimate rainfall totals compared to the observations, the offset between the two is fairly consistent, so for examining trends in monsoon rainfall the model should be adaquate.

It is interesting to note that the model used here can reproduce the main features of the EASM, including the Meiyu front and its progression. This has been challenging for models in the past including many CMIP5 generation models. 
The improved resolution of models from N96 (as used by most CMIP5 models) to N216 (as used by our model) has been shown to produce more realistsic precipitation globally \citep{Demory2014} and regionally \citep{Schiemann2014, Vellinga2016}, and more realistic monsoon systems \citep{Johnson2016}.
Our model uses prescribed SSTs and sea ice coverage, one advantage of which being that it will capture many ongoing large-scale modes of natural variability, such as El Nino. This and the `correct' forcing from sea surface temperature will allow a more accurate monsoon to be produced for a specific year than a coupled model. 
The physical realism of our model make it a suitable tool for studying changes in the characteristics of the EASM.

\section{Analysis of trends in monsoon rainfall}

We calculate anomalies with respect to the 1960-1979 mean value for the each of ALL and NAT and observations to illustrate trends in monsoon rainfall. Anomalies are only calculated for illustrative purposes, and do not inform the results shown below. We choose this baseline which is shorter than the more commonly used 1961-1990 baseline, in order that the reader might see changes in the metrics examined by eye.

The time series in Figure~\ref{tots_and_dry} shows the seasonal total monsoon rain anomaly for the SEC region, and we use SEC as an example for the rest of the results presented. No clear trend is seen for the time series of the seasonal total monsoon rain and the interannual variability is large for all of the regions indicated in Fig~\ref{obs_stations}. There is no clear difference between the ensemble means of the ALL and NAT forcings experiments for most of the timeseries, however there is a difference between the two for the most recent 5 years (2010-2015). 
The time series of mean daily total rainfall also shows no trend and large variability (not shown), and similarly variable time series, with lack of clear trends, are found for mean 5-day total rain, and maximum 5-day total rain. Since the time series data is very noisy, and trends are likely to be well within the internal variability, we focus on the differences between the distributions of the ALL and NAT ensembles for the most recent 15 years when presenting quantitative results.

Figure~\ref{tots_and_dry} also shows the total number of dry days in the monsoon season (rainfall total less than 1mm/day). For all regions the ALL forcings ensemble mean shows an increased number of dry days compared to the NAT ensemble mean, and the difference between the two ensembles appears greatest in more recent years suggesting an increasing trend in dry days in the ALL model. The variability of the model and the observations are again quite large and trends (if present) are not very clear. Given that the monsoon total rain shows no clear change, an increase in the number of dry days during the monsoon could imply an increase in rainfall total per day on wet days. 

Previous studies have noted changes in observed rainfall when the season is divided up into deciles of daily total rain \citep[e.g. ][]{Liu05,FuDan13,Fu08}. For our model ensemble we divide all the wet days (total rain $\ge$ 1 mm/day) in the monsoon season into deciles of daily total rain - where each decile contains 10\% of the total seasonal rainfall. We define the decile bin edges using all the members of the NAT ensemble between 1960-2015. The upper and lower limits for each bin are then applied to the ALL forcings ensemble. Figure~\ref{deciles_fig} shows the change in total rain in each decile for the last 20 years of data with respect to the 1960-1979 baseline climatology. Some regions show changes in the distribution of rainfall totals between deciles for the ALL ensemble mean. For the southern regions a clear increase can be seen in the lowest decile (bottom 10 \% daily total rain), and at the same time a decrease in the total rain in the upper deciles for the ALL ensemble with respect to the NAT ensemble (see Fig~\ref{deciles_fig}).
A decrease in rainfall from upper decile days and an increase in rainfall from lower decile days is the opposite of what is generally reported in the literature \citep[e.g. ][]{Liu05, Ma2017}, however the literature reports results for observations which end in 2000-2006. 

We also analyse the distribution (PDF) of daily totals (and numbers of days) within the 1st and 10th deciles (ie, the top and bottom 10\% daily total rainfall). For the 10th decile, comparing the ALL and NAT forcings experiments, the ALL ensemble has lower total rainfall and a lower number of days of rain, however the PDF (Fig~\ref{deciles_fig}) also shows a fatter tail at high values of mean rainfall per day. So even though the total rainfall in the 10th decile is less in the ALL ensemble than the NAT ensemble, the total rain in individual days is shifted to higher values (see Fig~\ref{deciles_fig}). We discuss this further below.

In reality rain falls during storms, which may last several days. We divide the monsoon season up into storms, or events, of n-days in duration. An event is defined as a number of consecutive days where each day has total rainfall greater than 1 mm. The duration of an event is {\bf n\_days}, the total rain which falls during an event is {\bf n\_day\_tot}, and the mean rainfall per day during an event is {\bf intens} \citep[see][]{Burke2016}. We divide up the monsoon into events for each grid cell. 

In a time series of mean and maximum annual n\_days, n\_day\_tot and intens (not shown) there is no clear trend, no clear separation between ALL and NAT ensemble means and large variability. As illustrated above, changes in monsoon rainfall are more pronounced at the extreme light and heavy ends.
In our previous paper \citep{Burke2016} we found that for rainfall events in May 2015 with high n\_day\_tot, intens increases and n\_days decreases in the ALL forcings ensemble compared to NAT. We examine the changes in n\_days and intens for the 95th percentile n\_day\_tot, where the 95th percentile is defined from the NAT ensemble for events between 1960-1979. Figure~\ref{ndays_intens_fig} shows the time series (percent anomaly) of n\_days and intens for events in the 95th percentile of n\_day\_tot - both figures show 5 year means in order to show the signal more clearly without so much natural variability. In this figure a trend can be seen for increased intens and decreased n\_days with time, and a shift in the spread of the ALL ensemble in the same direction.

We remind the reader that our chosen threshold for a wet or rainy day is 1 mm/day. Given that this threshold for a rainy day is set relatively low, this will inevitably lead to us recording long duration events using our n\_days method. The most extreme consequence of this being that our rainfall events can last weeks; a continuous rainfall event of this magnitude would probably be unphysical in reality. Given the temporal resolution of data available to us we are not able to examine the `real' duration of individual rain storms. However, the number of consecutive days of rain is an interesting metric with regards to flooding. The change in number of consecutive days of rain and the total rainfall in those days is also informative as to how the nature of rainfall during the monsoon season is changing as a result of anthropogenic forcings. As the EASM season progresses, the rain band (Meiyu front) moves northwards across East China and later retreats southwards again (as described in the introduction). As such most regions of East China will experience multiple wet and dry spells throughout the season. Our n\_days method allows us to see how anthropogenic forcings change in the progression and duration of the wet and dry spells.

\section{Change in likelihoods of extreme rainfall due to anthropogenic climate change}

We examine change in likelihood of the metrics for which we can see differences between the ALL and NAT experiment output described above using the most recent 20 years of model data (1996-2015). The change in probability, $\Delta$P (sometimes refered to as `risk ratio' in the literature), is given by $\Delta$P = P(ALL)/P(NAT), where P(ALL) and P(NAT) are the probability of a metric exceeding a given threshold in the ALL and NAT ensembles respectively. For each metric presented we define a threshold based on the NAT ensemble - these thresholds are the mean, 10th percentile or 90th percentile of the NAT ensemble depending on the metric examined. As such P(NAT) will be equal to 0.5 where we define our threshold to be the mean of NAT (etc). 

\begin{figure}
\centering
\includegraphics[width=15cm]{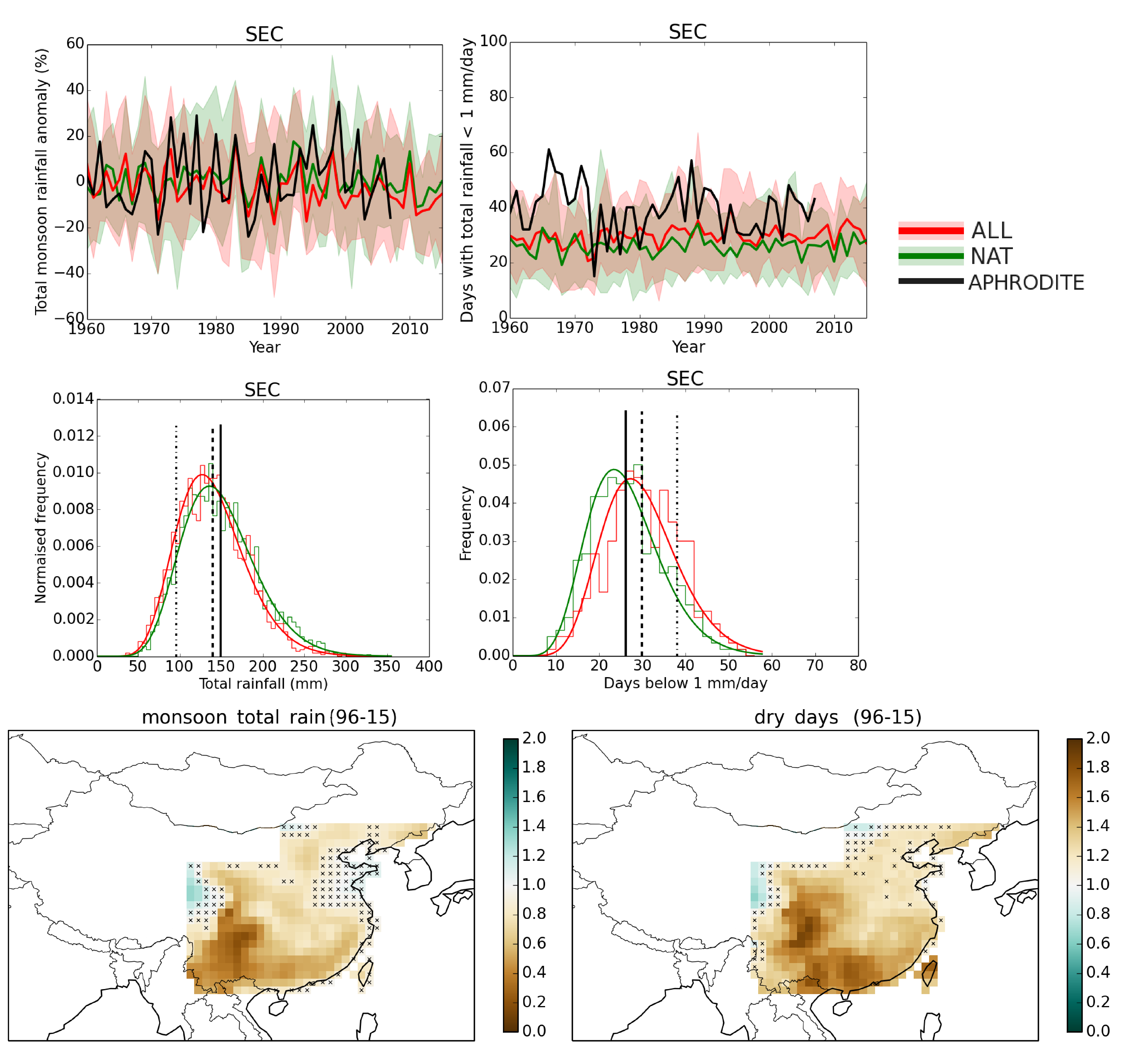}
\caption{Top row: time series for SEC of monsoon season total rainfall (left, anomaly with respect to 1960-1979) and total dry days during the monsoon (right). Colours as Fig~\ref{mons_shape_fig}.
Middle: Histograms with fitted PDFs for the most recent 20 years of the time series (1996-2015) for ALL and NAT, black line indicates the mean of the NAT ensemble, dashed line indicates the mean of ALL ensemble, dot-dashed line indicates 10th and 90th percentile of NAT ensemble for total rainfall and days below 1 mm respectively.
Bottom: Probability ratio ($\Delta$P) maps between ALL and NAT ensembles, with respect to the mean of the NAT ensemble for all ensemble members between 1996-2015. Black crosses indicate grid cell where $\Delta$P is not significant at a 2$\sigma$ (95 percent) level.}
\label{tots_and_dry}
\end{figure}

\begin{figure}
\centering
\includegraphics[width=15cm]{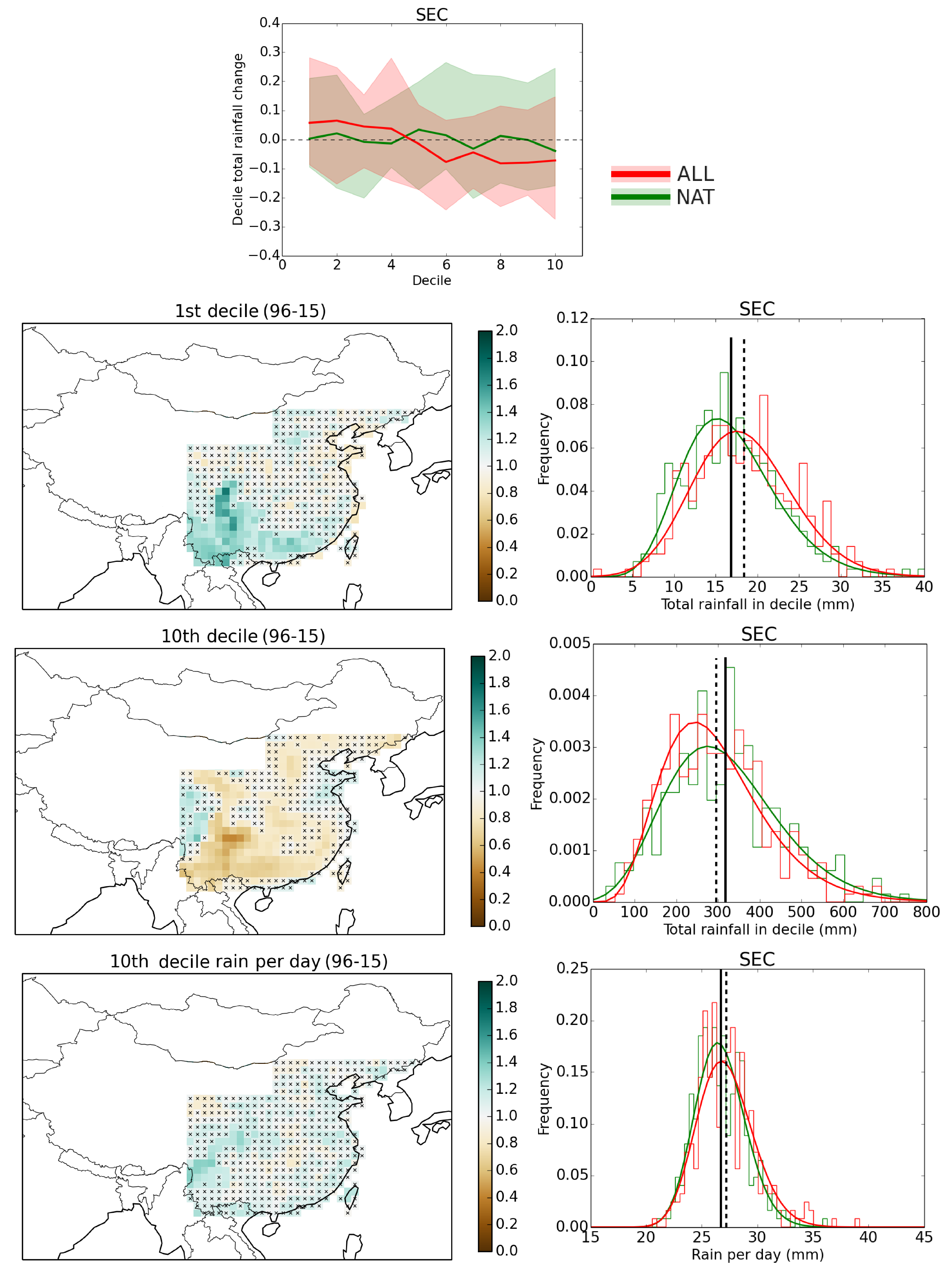}
\caption{Top: Fractional total rainfall change for 1996-2015 compared to 1960-1979 for each decile of daily rainfall for SEC.
Left column: $\Delta$P maps for 1st decile total rain (top), 10th decile total rain (middle) and 10th decile rain per day (bottom),  with respect to the mean of the NAT ensemble for all ensemble members between 1996-2015. Black crosses indicate grid cell where $\Delta$P is not significant at a 2$\sigma$ (95 percent) level. 
Right column: Histograms for the variables in the maps shown for SEC. Solid line indicates the mean of NAT, dashed line indicates the mean of ALL.}
\label{deciles_fig}
\end{figure}

\begin{figure}
\centering
\includegraphics[width=15cm]{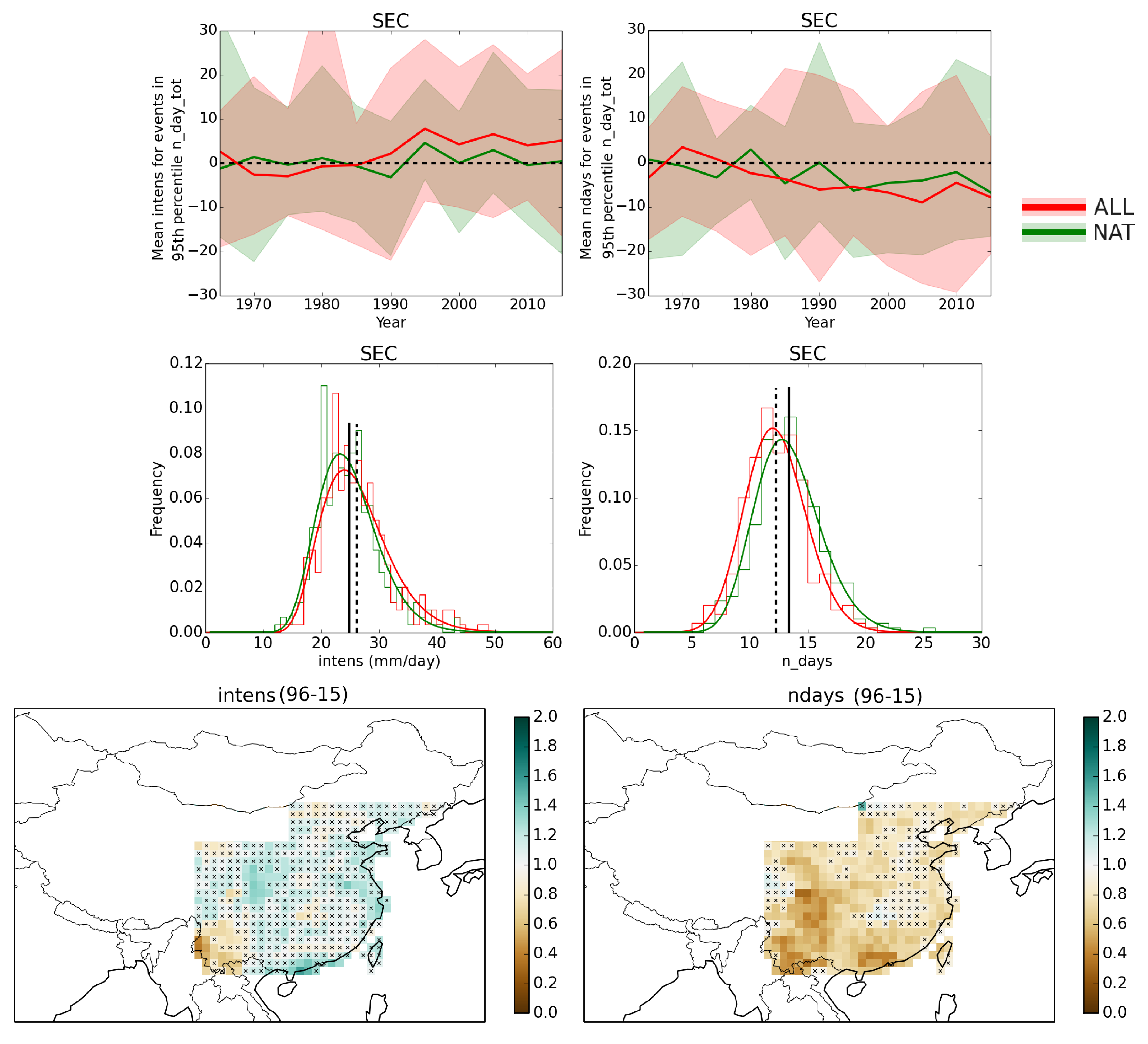}
\caption{Top: time series of intens (left, mm/day) and n\_days (right) for 95th percentile n\_day\_tot events for SEC.
Middle:Histograms with fitted PDFs for variables examined for all events between 1996-2015 in ALL and NAT.
Bottom: $\Delta$P maps, with respect to the mean of the NAT ensemble, for all ensemble members between 1996-2015, for events in the 95th percentile (w.r.t. 1960-1979) of n\_day\_tot. Black crosses indicate grid cells where $\Delta$P is not significant at a 2$\sigma$ (95 percent) level. } 
\label{ndays_intens_fig}
\end{figure}

P(ALL) is calculated by fitting a probability distribution function to the histogram of the variable considered, and taking the area under the curve above (or below) the threshold defined by NAT. This is illustrated in the PDF plots in Figures~\ref{tots_and_dry}--\ref{ndays_intens_fig}. We fit a gamma distribution to the normalized histogram for the variable considered, (as illustrated in the figures) using a maximum-likelihood estimation fitting routine (gamma.fit - freely available in scipy.stats). There are a minimum of 300 data points in each fitted histogram (15 members x 20 years x points per year for metric in question), so there is sufficient data for a reliable fit - by eye the curves appear to fit well. We test the goodness of fit by calculating $\Delta$P from the area under each histogram before fitting, and compare with the value of $\Delta$P from the fits to the histograms. We find the values of $\Delta$P from the histogram to be the same as those from the gamma fit to within 2\% (ie $\Delta$P(gamma fit)/ $\Delta$P(histogram)=1.00$\pm$0.02, SD=0.05). The results from calculating $\Delta$P with and without fitting are close enough that we are confident of the appropriateness of the gamma fit to represent the distribution of the data. These values derived with and without fitting are similar enough, and enough data is available to sample the distribution of values well, that fitting may not actually be necessary for examining extremes in this case.

The maps in figs~\ref{tots_and_dry}--\ref{ndays_intens_fig} also indicate which grid cells have $\Delta$P which is significant at the 2$\sigma$ level. The statistical significance of $\Delta$P is determined by bootstrapping the data and fitting the resulting histogram with a PDF from which $\Delta$P is calculated. The bootstrap is performed 1000 times for each grid cell (with replacement). For some of the figures there are a large number of grid cells which aren't significant at 2$\sigma$, and at a 1$\sigma$ level the picture is generally the same but with the addition of the grid cells along the coastlines also being significant. However, given the contiguous large areas showing similar changes in distribution, a lack of statistical significance in individual grid cells may be indicative of the presence of weak trends. We report area mean values for $\Delta$P and the change in the mean absolute value (also 10th and 90th percentile for monsoon total rain and number of dry days respectively) for each variable and each region in Table~\ref{results_tab}. The change in mean absolute value is defined as the difference between the mean of the NAT and the mean of the ALL ensembles (similarly for the value of 10th and 90th percentiles). As is clear in the table, when averaged over larger areas the values of $\Delta$P and the changes in absolute values of variables measured are indeed statistically significant in most cases.

Figure~\ref{tots_and_dry} shows $\Delta$P maps for the monsoon total rain and the number of dry days during the monsoon. Despite no clear difference between ensemble means and no clear trends being seen in the time series, the change in the probability distribution function of monsoon total rain between the ALL and NAT forcings ensembles is statistically significant (see also Table~\ref{results_tab}). Over all of East China the seasonal total rain is likely to be less, and the number of dry days during the monsoon is likely to be greater in the ALL ensemble compared to the NAT ensemble.
The total rainfall during the monsoon season is 10--40\% ($\Delta$P = 1.1--1.67) more likely to be below the NAT ensemble average in the ALL ensemble than the NAT ensemble. This is more severe in the south of the region of China examined than the north, see figure~\ref{tots_and_dry}. The area-mean value of total monsoon rainfall is found to be 45mm less in the ALL ensemble compare to NAT. The decrease in mean annual rainfall ranges from tens of mm in northeast China to $\sim$100 mm or more in southern areas (the maximum decrease for an individual grid cell examined is 291mm). The area-average $\Delta$P for total monsoon rainfall to be below the 10th percentile defined by NAT is 1.1, and the value of the 10th percentile seasonal total is decreased by 49mm in the ALL-forcings world compared to the NAT-forcings world (East China area average).

Similarly the likelihood of the number of dry days in the season being above the NAT average is $\Delta$P=1.4--2 in most of southern and eastern China, with an increase in the mean number of dry days of 3.6 days in the ALL ensemble. The area-mean likelihood of the number of dry days exceeding the 90th percentile of the NAT ensemble is $\Delta$P=1.9 in a world with climate change, with the 90th percentile number of dry days increased by 3.4 days in the ALL-forcings ensemble.

Figure~\ref{deciles_fig} shows $\Delta$P maps for bottom 10\% and top 10\% daily rainfall totals (first and tenth deciles), and the mean rainfall per day in the top 10\%. On average, there is likely to be more rainfall in the first decile and less rainfall in the 10th decile in the ALL ensemble compared to NAT. However the rainfall total on individual days in the 10th decile is likely to be greater in the ALL-forcings world - whilst this change is not statistically significant for the majority of individual grid cells, it is statistically significant when we average over larger areas (see Table~\ref{results_tab}). The strongest results for this are in South East China - for the total rain in the 1st decile being above the NAT average $\Delta$P=1.2, and for mean total rain in the tenth decile being below the NAT average $\Delta$P=1.25. However the likelihood of rainfall per day in the 10th decile being above the NAT mean in this area is $\Delta$P=1.1 in the ALL ensemble. So in this region, anthropogenic forcings may be causing shift to more light rain and less heavy rain in the season, but even though heavy rain days are more infrequent, the total rainfall per day on heavy rain days is increased. The likelihood changes we find for the number of days in each decile are similar in value to those reported above for total rain per decile. However the absolute changes in number of days in each decile are of the order 0.1-0.5 days increase, or 0.5-1.0 days decrease, for first and tenth deciles respectively. It could be argued that over the period of time examined, 1960--2015, this change is small enough to not be observable.

Figure~\ref{ndays_intens_fig} shows $\Delta$P maps for the duration (n\_days) and intensity (intens) of rainfall events in the 95th percentile of n\_day\_tot. For a NAT-forcings world average 95th percentile n\_day\_tot event, in an ALL-forcings world the event is 1.3 times (area average) more likely to be shorter in duration, and the daily total rain within each day of the event is 1.1 times more likely to be greater. On area-average, these events will be 1.8 days shorter, with the decrease in duration being more pronounced in the south than the north (see figure). The mean rain per day in these extreme events is increased by 1 mm/day in the ALL ensemble compared to NAT. Thus we have found evidence that the intensity of the most extreme rainfall events is expected to increase due to anthropogenic forcings.

\section{Discussion}
Under anthropogenic forcings the model predicts that there is, on average, a decrease in the total monsoon rainfall, an increase in the number of dry days, an increase in the total rain which falls in the 1st decile of daily totals and a decrease in the total rainfall in the 10th decile of daily total rain. This gives a picture of a generally dryer monsoon. However, for extreme heavy rainfall events a different picture is given. The results show an increase in total rain per day in the 10th decile of daily total rain, and for the 95th percentile of n-day-total rainfall in events as defined above, the mean rainfall per day is increased and the number of days over which the rain falls is decreased. So whilst the total seasonal rain is generally reduced, and the distribution of daily total rain is shifted towards the lighter end, for heavy rain events the rainfall per day is increased and the duration of heavy rain events is decreased.

The statistical significance of the changes reported per grid cell is strong for the general drying changes - monsoon total rainfall, number of dry days, increase in 1st decile days, shortening in duration of extreme events. The statistical significance per grid cell is weaker for increased tenth decile rain per day and increased intensity of heavy rain events. Figures~\ref{deciles_fig} and \ref{ndays_intens_fig} show comparatively few grid cells are significant at 2$\sigma$ for these metrics compared to the drying metrics (at 1$\sigma$ the coastal grid boxes also appear significant, but otherwise the figures are very similar, not shown). However, for regional averages on most metrics the results are statistically significant (see Table~\ref{results_tab}). The heavy rainfall changes are smaller in magnitude compared to the changes for drying metrics for both grid cells and regional means. This suggests that the increase in extremes is a smaller effect than the overall drying. 

We have examined changes in the monsoon season, considering all the days from the beginning of April to the end of August as being in the season. As illustrated in figure~\ref{mons_shape_fig}, the rainfall within the season is very variable between dates and locations. It may be that our examination misses detail on shorter timescales and that changes in extremes are more or less pronounced on the monthly timescale than that reported for the whole season. We also do not examine changes in the timing or spatial extent of the monsoon season.

We point out that our results are for model data and represent changes in likelihoods between model ensembles with and without anthropogenic climate change. As such the results presented here are predictions of the changes in monsoon rainfall as a result of anthropogenic forcing, which we might expect to see in observations.

Whilst we have carried out some verification with the observations available to us, we suspect that the observations we have for this region are imperfect (as illustrated in figure~\ref{mons_shape_ver}). In order to verify the model and results presented here more detailed and up to date observational studies will be required. Unlike many CMIP5 generation models which struggle to reproduce extreme rainfall observed in reality, the model set-up used is able to produce the extremes of rainfall which are observed, and tends to over rather than under predict the extremity and frequency of heavy rainfall (however the observed gridded data we compare to may underestimate extreme rainfall).

\subsection*{Physical basis and comparison with previous studies}

In recent years there have been reports of a southern flood / northern drought pattern during the summer monsoon (see introduction). A dryer monsoon season could easily lead to drought, and short intense rainfall bursts can lead to flooding. Long duration rainfall is generally needed to alleviate droughts, so short but heavy rainfall events, once over, may allow a drought to persist. Examination of the mechanism which would cause extended drought over northern China but recurring flooding over southern China is outside of the scope of this study. 

Several previous model-based studies discuss intensifying convection as a result of global warming leading to increased heavy rainfall, and depletion of light rain at the expense of this heavy rain \citep[e.g.][]{Trenbeth2003}. The proposed mechanisms for this change are that global warming can lead to enhanced convection processes, an enhanced water cycle and increased convergence (super Clausius-Clapeyron). The heavy rainfall as the result of these processes is more extreme than in a world without anthropogenic climate change, and the result of intense downpours is that the precipitable water column is emptied, inhibiting subsequent light rainfall \citep{Fowler95, FisherKnutti16,OGormanSchneider}. The recent observational work of \cite{FisherKnutti16} shows that globally very heavy daily total rainfall events in the 95th percentile or greater are notably increasing in frequency and this is reflected in current climate models. Generally, recent observational studies of global rainfall trends report a slight increase in total rainfall \citep[e.g.][]{Wu13}, however for heavy rainfall a significant increasing trend is consistently found \citep{Donat16, Westra13, Ban15, OGormanSchneider}. 

Over the area of East China, in the upper decile of daily rainfall total we see some weak shift to larger rainfall per day values, but we do not see a reduction in light rain (1st--2nd decile daily total rain). Perhaps by selecting the 90th percentile, rather than the 95th or 99th we are only seeing hints of this trend in our only moderate results for heavy rain increase. Similarly for our 95th percentile n-day-total rainfall, we see some weak indication of increased daily total, but it is not as impressive as that reported for global daily totals.

On more local spatial scales, some previous observational studies also report an increase in heavy rain and a decrease in light rain over China. For example \cite{Ma2017}, observe a decrease in total rain from light rain days, and an increase in total rain from heavy rain days. Their reported change in light rain is weak statistically, and their reported change in heavy rain is larger and statistically stronger.

Numerous observational studies have reported an increase in seasonal total rainfall over the period 1960-2000 for eastern China \citep{Liu05, Zhai04, WangZhou05, Su06, FuDan13, Fu08, QianQin07, Gemmer03}. However these changes are not uniformly spatially coherent, nor are the observed regions all defined to cover the same areas as each other, or as that examined here. Subsets of these works \citep{Zhai04,Liu05,FuDan13,Fu08, Su06, Qian2009} also report increases in the number of heavy rain days and decreases in light rain days, and also with shifts in rainfall totals across daily deciles in a similar direction. The method by which deciles or thresholds for extreme rainfall totals are defined differs between most of these stuides, being defined for individual seasons in some, and annually in others. The regions studied also vary between publications, and deciles and extremes may be defied as an area average or within sub-regions. Additionally, these studies tend to end in 2000, near the start of our current climatology and given that they end 15 years ago it would be interesting to see if the results that they present continue in more recent years. Similarly to the result presented here, the trends reported by most literature studies tend to be statistically weak and the data noisy - this is a frequent issue for studies of precipitation.

There are observational literature studies which are complementary to our findings. For example \cite{Xiao2016} examine the observed hourly peak total rainfall during the monsoon season. They find peak hourly rainfall is correlated with daily mean temperature, and that the number of rain hours per day decreases with increasing temperature, with hourly precipitation extremes increased by 10\% per degree increase of daily mean temperature. However they find daily extremes decrease by approximately the same amount - so extreme total rainfall is increasing but duration is decreased on the hourly timescale.

\cite{Liu05} find a 10\% decrease in frequency of precipitation events between 1960-2000. \cite{Zhai04} also report a decrease in number of rain days over East China between 1950-2000. They also find the daily rainfall total in the 95th percentile has increased with time, and an increased frequency of 95th percentile rainfall days in South and Eastern China during the warm half of year. However they find no statistically significant change in annual rainfall total.

Precipitation is a notoriously difficult variable to measure accurately, perform trend analysis of, and detect changes in with any meaningful confidence. In the studies discussed above, several subtly different methods are used to detect changes in rainfall in subtly, but non-trivially, different ways. In an ideal world it would be beneficial to have a unified metric, or set of metrics by which changes in rainfall could be judged. This would help promote a clearer path to detecting and attributing changes and understanding what drives them.

\subsection*{Future changes}

With future reductions in aerosol emissions and a continued increases in greenhouse gas emissions, historical trends in monsoon rainfall may not be indicative of future changes \citep{Christensen2013}. CMIP5 \citep{CMIP5} RCP8.5 model projections predict that east China summer season (JJA) will become wetter in future (see figure 12.22 in IPCC AR5 Chapter 12, \cite{IPCC}), with a projected increase of approximately 20\% in seasonal rainfall total by the end of the century with respect to the mean of 1986-2005. The projected changes are likely due to increases in GHGs and reduction in aerosols. Additionally, in line with our historical results, the maximum 5-day precipitation and the number of consecutive dry days are projected to continue to increase for east China \citep[see figure 12.26 of IPCC AR5 Chapter 12; also Chapter 14, page 1271,][]{Christensen2013}. 

In line with our results for historical changes in rainfall, in future, in a world with increased global warming, we might expect to see more short intense rainstorms, increasing the possibility of flash flooding. However, there may be fewer days of rain between extreme rainstorms, which can lead to drought. Alleviation of drought requires rain over an extended period, the shortening of rainstorms means that drought may be exacerbated.

\section{Conclusions}
We have presented the results of a historical model ensemble with and without anthropogenic influence on the climate system. We verify our model against observed climatology and find that it can reproduce the main features of the EASM. The model shows that, in the anthropogenic influence scenario, the EASM is generally dryer overall, with a decrease in total rain and an increase in dry days. However the anthropogenic influence model also shows an increase in the intensity of heavy rain events. These changes could lead to increased likelihood of flash flooding during rainstorms, but also an increased likelihood or severity of drought in some locations. 

Historically a range of different results are found when exmaining observed rainfall in Eastern China during the summer and EASM season. These changes are not always consistent with those observed gloablly, which suggests localised forcings may be at play. However, given the range of methodologies and obeserved and modelled data available for investigating rainfall, this is an area which still warrants further study.


%
\acknowledgments
This work was supported by the UK-China Research \& Innovation Partnership Fund through the Met Office Climate Science for Service Partnership (CSSP) China as part of the Newton Fund and by the Joint DECC/Defra Met Office Hadley Centre Climate Programme (GA01101).

%






%
%
%
\bibliographystyle{ametsoc2014}
\bibliography{references}

%

\begin{table}[t]
\caption{Results by regions as indicated in Figure~\ref{obs_stations}. Probability ratio, $\Delta$P, values give the change in likelihood of the mean seasonal value of the variable considered for the ALL ensemble with respect to the NAT ensemble. The absolute change is the change in the value of the variable considered for ALL ensemble with respect to the NAT ensemble, for example, the mean seasonal rainfall total is X mm less. Results {\bf not} statistically significant at 2$\sigma$ are highlighted in {\it italics}.}
\label{results_tab}
\begin{center}
\begin{tabular}{llccccccc}
\hline\hline
Variable & & NE & NCC & NEC & CW & SCC & SEC & SE \\
\hline

\hline

Total rainfall          &  $\Delta$P mean               & 0.8$\pm$0.01     & 0.8$\pm$ 0.01  & 0.9$\pm$0.01   & 0.9$\pm$0.03    & 0.6$\pm$0.03    & 0.8$\pm$0.01   & 0.6$\pm$0.02  \\
                        &  Mean change (mm)             & -28.3$\pm$1.72  & -34.9$\pm$2.26 & -13.5$\pm$2.38 & -13.3$\pm$4.62  & -110.0$\pm$6.56 & -72.9$\pm$4.12 & -146.2$\pm$9.15 \\
                        &  $\Delta$P 10th percentile    & 0.9$\pm$0.01    & 0.9$\pm$0.01   & 1.0$\pm$0.01   & 1.0$\pm$0.01    & 0.8$\pm$0.01    & 0.9$\pm$0.01   & 0.9$\pm$0.01   \\
                        &  Mean change (mm)             & -23.4$\pm$1.82  & -33.9$\pm$2.93 & -14.5$\pm$2.49 & -13.3$\pm$4.98  & -98.3$\pm$6.44 & -48.65$\pm$3.27 & -108.1$\pm$7.87 \\

\hline
Dry days                &  $\Delta$P mean               & 1.2$\pm$0.01    & 1.3$\pm$0.02   & 1.2$\pm$0.01   & 1.1$\pm$0.04    & 1.5$\pm$0.02    & 1.4$\pm$0.01   & 1.6$\pm$0.01 \\
                        &  Mean change (days)           & 2.0$\pm$0.13    & 3.6$\pm$0.25   & 2.1$\pm$0.08   & 1.0$\pm$0.31    & 5.9$\pm$0.28    & 4.1$\pm$0.13   & 6.5$\pm$0.11 \\
                        &  $\Delta$P 90th percentile    & 1.4$\pm$0.03    & 1.9$\pm$0.07   & 1.5$\pm$0.04   & 1.4$\pm$0.10    & 2.6$\pm$0.11    & 1.8$\pm$0.03   & 2.5$\pm$0.06 \\
                        &  Mean change (days)           & 2.0$\pm$0.20    & 3.6$\pm$0.28   & 2.0$\pm$0.16   & 0.9$\pm$0.36    & 5.7$\pm$0.35    & 3.8$\pm$0.16   & 6.0$\pm$0.18 \\
  
\hline
First decile            &  $\Delta$P mean               & 1.0$\pm$0.01   & 1.0$\pm$0.01   & 1.0 $\pm$0.01  & 1.1$\pm$0.02    & 1.1$\pm$0.02    & 1.0 $\pm$0.01  & 1.2$\pm$0.02 \\
 total rain             &  Mean change (mm)             & 0.1$\pm$0.03   & {\it0.1$\pm$0.06}   & -0.1$\pm$0.03  & 0.5$\pm$0.18    & 0.7$\pm$0.13    & 0.2$\pm$0.06   & 1.0$\pm$0.12 \\

Tenth decile            &  $\Delta$P mean               & 0.9$\pm$0.01   & 0.9$\pm$0.01   & 1.0 $\pm$0.01  & 1.0$\pm$0.02    & 0.8 $\pm$0.02   & 0.9$\pm$0.01   & 0.8$\pm$0.02   \\
  total rain            &  Mean change (mm)             & -12.1$\pm$0.97 & -10.9$\pm$1.33 & {\it1.2$\pm$2.00}   & {\it-1.8$\pm$2.17}   & -44.4$\pm$3.72  & -25.7$\pm$2.44 & -52.5$\pm$5.75 \\

Tenth decile            &  $\Delta$P mean             & 1.0$\pm$0.01   & 1.1$\pm$0.01   & 1.1$\pm$0.01   & 1.0$\pm$0.01   & 1.0 $\pm$0.01   & 1.0$\pm$0.01   & 1.1$\pm$0.02 \\
 rain per day           &  Mean change (mm/day)         & 0.4$\pm$0.04   & 0.9$\pm$0.03   & 0.9$\pm$0.13 & 0.1$\pm$0.03    & 0.3$\pm$0.07    & 0.5$\pm$0.09   & 0.6$\pm$0.09 \\

\hline
n\_days                 &  $\Delta$P mean               & 0.8$\pm$0.01   & 0.8$\pm$0.01  & 0.8$\pm$0.01   & 0.8$\pm$0.02    & 0.7$\pm$0.02    & 0.8 $\pm$0.01  & 0.6$\pm$0.01  \\
                        &  Mean change (days)           & -0.6$\pm$0.04  & -0.9$\pm$0.09  & -0.6$\pm$0.04  & -1.9$\pm$0.46   & -2.6$\pm$0.21   & -1.1 $\pm$0.13 & -4.9$\pm$0.29 \\

intens                  &  $\Delta$P mean               & 1.0$\pm$0.01   & 1.1$\pm$0.01  & 1.1$\pm$0.01   & 1.0$\pm$0.02    & 1.1 $\pm$0.02   & 1.1 $\pm$0.01  & 1.1$\pm$0.02 \\
                        &  Mean change (mm/day)         & 0.4$\pm$0.12   & 0.9$\pm$0.10  & 2.2$\pm$0.20   & 0.1$\pm$0.04    & 0.7$\pm$0.14    & 1.3  $\pm$0.21 & 1.4$\pm$0.2 \\

\hline
\end{tabular}
\end{center}
\end{table}

%

\end{document}